\documentclass[twoside,12pt]{article}
\usepackage{epsfig}
\begin{document}

%
%
\title{ \vspace{1cm}
 Virtual Compton Scattering at Low Energy and the 
Generalized Polarizabilities of the nucleon}

\author{H.Fonvieille \\
Laboratoire de Physique Corpusculaire IN2P3-CNRS \\
 University Blaise Pascal Clermont-II, 63177 AUBIERE, France \\
\ \\
}

\maketitle

%
\begin{abstract}

Virtual Compton Scattering on the nucleon: 
\ $\gamma^* N \to \gamma N$ \
is a new and rapidly developing field at low and high energies,
with the emergence of recent concepts such as Generalized Parton 
Distributions (GPDs) at high energy, and Generalized Polarizabilities 
(GPs) at low energy.
This lecture is about the low energy part, i.e. for energies
in the $( \gamma p )$ center-of-mass mainly up to 
the $\Delta(1232)$ resonance region. I review the concept of 
GPs and the experiments dedicated to their measurement.

\end{abstract}

\section{The Generalized Polarizabilities of the Nucleon}

The formalism of Polarizabilities in Real Compton Scattering (RCS) and
Generalized Polarizabilities in Virtual Compton Scattering (VCS) 
has been the subject of previous lectures at the Erice School of Nuclear 
Physics by Nicole D'Hose.
I refer the reader to her very detailed and pedagogical lecture 
on the subject~\cite{ericedhose}. I will summarize the  concepts again,
and will concentrate on some recent experimental developments 
in the field.

\subsection{The physical meaning of Generalized Polarizabilities} 
\label{secmeaning}

The GPs are new observables to study nucleon structure. They have 
a natural connection with two well-known processes:
RCS and electron-nucleon elastic scattering.

\begin{figure}[ht]
\begin{center}
\begin{minipage}[t]{12 cm}
\epsfig{file=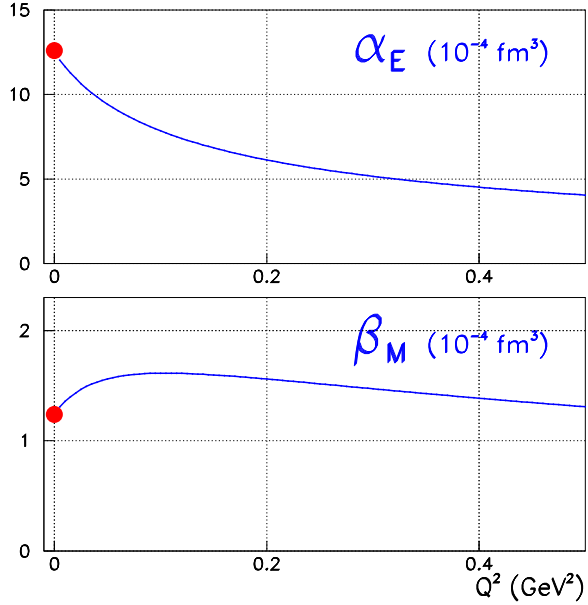,scale=0.45}
\end{minipage}
\begin{minipage}[t]{16.5 cm}
\caption{The electric and magnetic GPs of the proton 
as calculated in heavy baryon chiral perturbation theory~\cite{chptop3}. 
The magnitude ($10^{-4}$ fm$^3$) reminds us that the
nucleon is a strongly bound and very rigid object.}
\label{fig01}
\end{minipage}
\end{center}
\end{figure}

\vskip 2mm
{\it 1. From  $Q^2$=0 to finite $Q^2$ :} 
when going from RCS : \ $\gamma N \to \gamma N$ \ to
VCS : \ $\gamma^* N \to \gamma N$ at photon 
virtuality $Q^2$, the
electromagnetic polarizabilities of the nucleon become
functions of  $Q^2$ and are called the Generalized Polarizabilities.
This concept was first introduced in 1995~\cite{gui95} and allows
to probe the polarizability locally inside the nucleon, with a 
distance scale given by $Q^2$.
For example, in most models the electric GP \ $\alpha_E$ \ is predicted to 
decrease monotonically with $Q^2$, while the magnetic GP \ $\beta_M$ \ 
is predicted to go through a maximum
before decreasing, as shown in figure~\ref{fig01}. 
This last feature is usually explained 
by the dominance  of diamagnetism due to the pion cloud 
at long distance, or small $Q^2$,  
and the dominance  of paramagnetism due to a quark core 
at short distance, or large $Q^2$.
In VCS there are also new GPs due to the longitudinal polarization
state of the virtual photon.

\vskip 2mm
{\it 2. From finite q' to q'=0 :}
similarly to the polarizabilities in RCS, the GPs in VCS are theoretically 
defined in the limit q'$\to$0, where q' is the energy of 
the final real photon. This is the ``zero-frequency'' limit
corresponding to a static electromagnetic field.
In this limit, from kinematic point of view the VCS process 
 $\gamma^* N \to \gamma N$ becomes the process $\gamma^* N \to N$, 
i.e. elastic electron-nucleon scattering. 
As stated in ref.~\cite{guivdh98}, {\it ``VCS at threshold can 
be interpreted as electron scattering by a target which is in constant 
electric and magnetic fields''}.
The GPs can then be seen as Fourier transform of densities of 
electric charges and magnetization of a nucleon deformed by an 
applied EM field.

\subsection{The VCS amplitude and its multipole decomposition}

\begin{figure}[ht]
\begin{center}
\begin{minipage}[t]{12 cm}
\epsfig{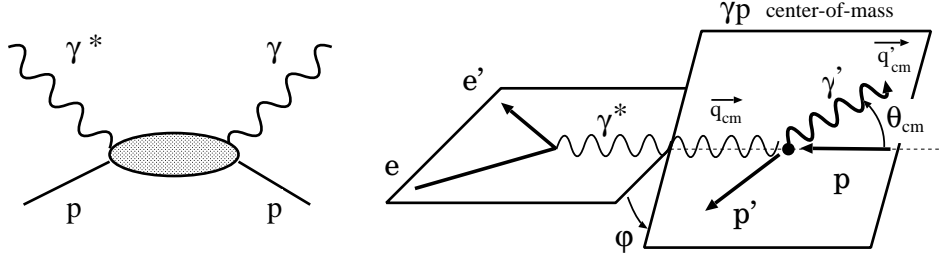}
\end{minipage}
\begin{minipage}[t]{16.5 cm}
\caption{The VCS graph on the proton and
the main kinematic variables of the process.
\label{fig02}}
\end{minipage}
\end{center}
\end{figure}

VCS is accessed experimentally by exclusive photon electroproduction 
as shown in Figure~\ref{fig02}. The main kinematic variables are
the CM three-momenta \ $q_{cm}$ \ and \ $q'_{cm}$ \ of 
the initial and final photons, and the CM angles 
of the outgoing real photon w.r.t. $\vec q_{cm}$ : 
the polar angle \ $\theta_{CM}$ and azimuthal angle $\varphi$
(see fig.~\ref{fig02}). Other useful variables are 
the virtual photon four-momentum transfer squared $Q^2$ 
and its polarization rate $\epsilon$.

\begin{figure}[ht]
\begin{center}
\begin{minipage}[t]{13 cm}
\epsfig{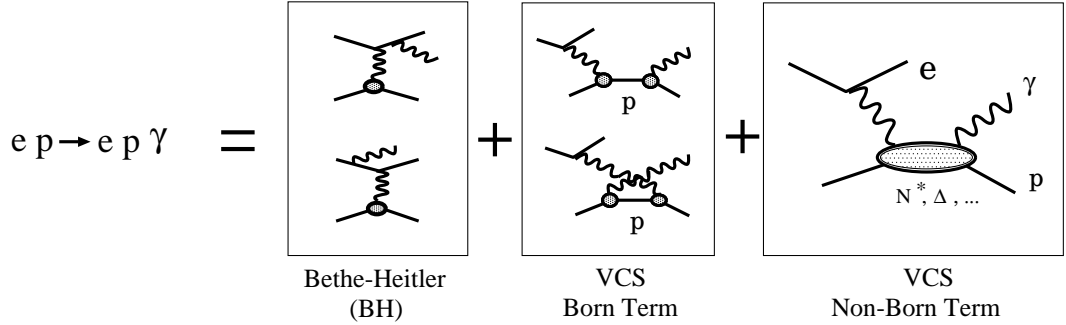}
\end{minipage}
\begin{minipage}[t]{16.5 cm}
\caption{Decomposition of the  photon electroproduction amplitude.
\label{fig03}}
\end{minipage}
\end{center}
\end{figure}

Due to electron scattering, one also has the Bethe-Heitler process (BH)
where the final photon is emitted by the incoming or outgoing electron.
Figure \ref{fig03} shows the decomposition of the photon electroproduction 
amplitude \ $T$ \ into the coherent sum of the BH, Born and Non-Born parts:

\begin{eqnarray}
T_{ep \to ep \gamma} \ = \ T_{BH} \ + \ T_{VCS \ Born} \ + \
 T_{VCS \ Non-Born}
\label{eq00}
\end{eqnarray}

The (BH) and (VCS Born) parts are known and entirely calculable, with the 
nucleon EM form factors as inputs. The Non-Born amplitude contains 
the unknown part that one aims to measure. The interest is that
this part describes the excited spectrum of the nucleon in 
the intermediate state, the excitation and de-excitation 
being specifically of electromagnetic origin.
\footnote{
It should be noted that the definition of the Born and Non-Born terms 
of the VCS amplitude is not unique~\cite{drebnb}.
In ref.~\cite{gui95} the Non-Born part includes the $\pi^0$ $t$-channel 
exchange graph, and the Born terms include the antinucleon graph.
}

The GPs are introduced using the decomposition
of the VCS Non-Born amplitude~\cite{gui95} into multipoles corresponding
to well-defined initial and final EM transitions. A multipole \ $H$ \ 
is characterized by five quantum numbers:
\ $L,L'$ = (orbital) angular momentum of the initial and final photons;
\ $\rho , \rho'$ = their polarisation states ($\rho = 0,1,2$ \ for 
longitudinal, magnetic, electric); $S=1$ (resp.0) allows for nucleon 
spin-flip (resp. non spin-flip) in the process.
A polarizability \ $P$ \ is defined in the zero-frequency limit by:

\begin{eqnarray}
\mbox{GP} \ = \ P^{ \ (\rho ' L' , \rho L)S} \ (q_{cm}) 
\ \ \ \ \ \ \sim \ \ \ \ \ \ 
lim {\big \vert }_{q'_{cm} \to 0} \ \ \ {\bigg ( } \ 
{\displaystyle 1 \over \displaystyle q'^{L'}_{cm} } \ 
{\displaystyle 1 \over \displaystyle q^{L}_{cm} } \   
H^{(\rho ' L' , \rho L)S}_{NonBorn} \ (q_{cm},q'_{cm}) 
\ {\bigg ) }
\label{eq01}
\end{eqnarray}

The GPs are a function of $q_{cm}$, or equivalently $Q^2$.
\footnote{ This four-momentum transfer is actually the photon
virtuality when $q'_{cm} \to 0$, at fixed $q_{cm}$ and $\epsilon$ .
It is given by :
\  $\tilde Q^2 = 2 M_N \cdot ( \sqrt{  M_N^2 + q_{cm}^2  } - M_N)$ . 
One also defines the virtual photon CM energy in the
same limit : \  $\tilde q _0 = M_N - \sqrt{ M_N^2 + q_{cm}^2 }$ 
\cite{guivdh98}.
}
The lowest order is \ $L'=1$ , leading finally to six independent 
(dipole) GPs after application of nucleon crossing symmetry 
and charge conjugation invariance~\cite{dre6gp,drebnb}. The choice of 
the six GPs is not unique; a standard set is given in table \ref{tab01}. 
The two scalar GPs ($S=0$) are 
a generalization of the electric 
and magnetic polarizabilities \ $\alpha_E$ and $\beta_M$.
Among the four spin GPs ($S=1$) some have a correspondence in RCS.

\begin{table}
\begin{center}
\begin{minipage}[t]{16.5 cm}
\caption{The six lowest-order GPs. Notations use either 
the EM transitions or the five indices. $S=0$ (resp. 1) correspond to
scalar (resp. spin) GPs.
}
\label{tab01}
\end{minipage}
\begin{tabular}{|c|c|c|c|c|c|c|}
\hline
\ & \ & \ & \ & \ & \ &  \\
final $\gamma$ & initial $\gamma^*$ & S  
& $P^{(\rho ' L' , \rho L ) S } (q_{cm})$ & $P^{X \to Y}$ 
& RCS limit & resonances \\
\ & \ & \ & \ & \ &  ($Q^2=0$) & involved \\
\hline
E1 & C1 & 0 & $P^{(01,01)0}$ & $P^{C1 \to E1}$ & 
$ - {4 \pi \over e^2} \sqrt{2 \over 3} \  \alpha_E $ & $D_{13}, S_{11}$ \\
M1 & M1 & 0 & $P^{(11,11)0}$ & $P^{M1 \to M1}$ & 
$ - {4 \pi \over e^2} \sqrt{8 \over 3} \ \beta_M $ & $P_{33}, P_{11}$ \\
E1 & C1 & 1 & $P^{(01,01)1}$ & $P^{C1 \to E1}$ & 0 & $D_{13}, S_{11}$ \\
M1 & M1 & 1 & $P^{(11,11)1}$ & $P^{M1 \to M1}$ & 0 & $P_{33}, P_{11}$  \\
E1 & M2 & 1 & $P^{(01,12)1}$ & $P^{M2 \to E1}$ & 
$ - {4 \pi \over e^2} {\sqrt{2} \over 3} \ \gamma_3$ & $D_{13}$ \\
M1 & C2 & 1 & $P^{(11,02)1}$ & $P^{C2 \to M1}$ & 
$ - {4 \pi \over e^2} \sqrt{8 \over 27} \ ( \gamma_2 + \gamma_4)$ & $P_{33}$ \\
\hline
\end{tabular}
\end{center}
\end{table}

Most models of nucleon structure at low energy give predictions 
for these new observables: 
heavy baryon chiral perturbation theory~\cite{chptop3,chptop4},
linear sigma model~\cite{linearsigma}, 
effective lagrangian model~\cite{vdheff}, 
dispersion relation model~\cite{drmodel1,drmodel2}, 
non-relativistic constituent quark models~\cite{nrqcm1,nrqcm2},
etc.
A review of the theoretical predictions can be found in ref.~\cite{nrqcm2}. 
Ideally, one aims at measuring the whole set of the six GPs
as a function of $Q^2$ to make the most complete comparison with 
theory.

\subsection{The pion threshold and methods to extract the GPs}

\begin{figure}[ht]
\begin{center}
\begin{minipage}[t]{13 cm}
\epsfig{file=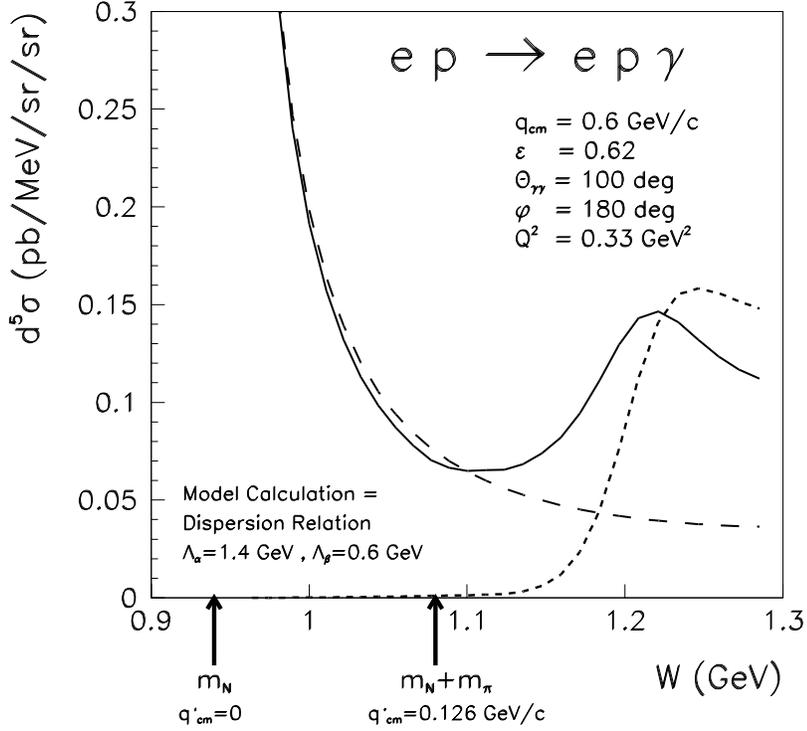,scale=0.6}
\end{minipage}
\begin{minipage}[t]{16.5 cm}
\caption{Example of photon electroproduction cross section
\ $d^5 \sigma / d k'_{lab} d \Omega _e' d \Omega_{pCM}$ \ 
($k'_{lab}$= scattered electron energy, $d \Omega _e'$ = scattered
electron lab solid angle, and $d \Omega_{pCM}$ = final proton CM 
solid angle). Contributions of BH+Born (long-dashed), Non-Born 
(short-dashed) and total cross section (solid).
\label{fig04}}
\end{minipage}
\end{center}
\end{figure}

The pion threshold is defined by \ $W= m_N + m_{\pi}$ \ where $W$ is the
total energy in the \ $(\gamma p)$ center-of-mass.
Above this threshold the VCS amplitude becomes complex. 
While $T_{BH}$ and $T_{VCS \ Born}$ remain real, 
the amplitude $T_{VCS \ Non-Born}$ acquires an imaginary part, 
due to the coupling to the $\pi N$ channel. 
Figure \ref{fig04} shows an example of  $(ep \to ep \gamma)$
cross section as a function of $W$. Below pion threshold, 
the effect of GPs, which is contained in $d \sigma_{Non-Born}$ , 
is small (10-15 \% maximum). It becomes important
in the region of the $\Delta (1232)$ resonance. It is worth noting 
that the GPs are defined at  $q'_{cm}$=0 but their contribution
to the cross section is zero at this point (see eq.~\ref{eq02}). 
Therefore measurements must be done at finite $q'_{cm}$.

There are presently two methods to extract GPs from measurements of
(unpolarized)  photon electroproduction cross sections. 
The first method uses a Low Energy Theorem (LET) and is valid below 
pion threshold only.
The second method is based on the Dispersion Relation (DR) approach
and its domain of validity includes the $\Delta(1232)$ resonance, 
up to the $N \pi \pi$ threshold.

\subsection{The Low Energy Theorem}

The Low Energy Theorem~\cite{low} was first applied to VCS 
by P.Guichon et al.~\cite{gui95}.
It leads to the following expression for the unpolarized 
$e p \rightarrow e p \gamma$ cross section below pion threshold :
\begin{eqnarray}
 d^5 \sigma _{EXP} &=&   
 d^5 \sigma _{BH+Born}  \ + \ {\big [ } \ \ q'_{cm} \phi \Psi_0 \ + \ 
{\cal O } (q'^2_{cm}) \ \ {\big ] } \ , \nonumber \\
\Psi_0 &=& v_1 \cdot 
(P_{LL} - {\displaystyle 1 \over \displaystyle \epsilon} P_{TT}) 
\ + \ v_2 \cdot  P_{LT}  
\label{eq02} 
\end{eqnarray} 
where $\phi, v_1, v_2$ are kinematic coefficients
defined in~\cite{guivdh98}.
The bracket on the right-hand side is a low-energy expansion (LEX)
in $q'_{cm}$. The unknown part of the nucleon structure
is contained in the $\Psi_0$ term (dipole GPs) and 
higher order terms (higher-order polarizabilities).
The three structure functions $P_{LL}, P_{TT}$ and $P_{LT}$ 
are the following combinations of the lowest-order GPs:

\begin{eqnarray}
\begin{array}{lll}
 P_{LL} (q_{cm}) & = & -2 \sqrt{6} \ M_N \ G_E \ 
 P^{(01,01)0}  (q_{cm}) \\
P_{TT} (q_{cm}) & = & -3 \ G_M \  
{ q_{cm}^2 \over {\tilde q _0} }  \times
{\bigg [ } \ 
 P^{(11,11)1} (q_{cm})  - \sqrt{2} \tilde q _0 
\  P^{(01,12)1}  (q_{cm})
\ {\bigg ] } \\
 P_{LT} (q_{cm}) & = & \sqrt{ {3 \over 2} } \ 
{ M_N \ q_{cm} \over {\tilde Q} } \ G_E \
 P^{(11,11)0}  (q_{cm}) +
{3 \over 2} \ { {\tilde Q} \ q_{cm} \over {\tilde q _0} } 
\ G_M \  P^{(01,01)1}  (q_{cm}) \\
\label{eq03}
\end{array}
\end{eqnarray}

The electric and magnetic form factors $G_E, G_M$ are taken at
four-momentum transfer squared $\tilde Q ^2$ (see footnote 2).
These equations tell us that \ $P_{LL}$ \ is proportional to 
the electric GP \ $\alpha_E$ ,  \ $P_{TT}$ \ is
a combination of two spin GPs, and \ $P_{LT}$ \ is a combination
of the magnetic GP \ $\beta_M$ \ and a spin GP.

In an experimental analysis based on the LET, one compares the
measured cross section \  $d^5 \sigma _{EXP}$ \ to the 
$d^5 \sigma _{BH+Born}$ cross section, entirely calculable from QED.
At $q'_{cm}$=0 these two cross sections coincide, as shown by 
eq.~\ref{eq02}.
\footnote{ In the term \ $q'_{cm} \phi$ \ multiplying $\Psi_0$ 
in eq.~\ref{eq02}, the factor $\phi$ remains finite when  
$q'_{cm} \to 0$. }
The \ $\Psi_0$ \ term is obtained by extrapolation of the
quantity \ 
$\Delta {\cal M} = 
 (d^5 \sigma _{EXP} -  d^5 \sigma _{BH+Born})/( q'_{cm} \phi )$
\ to \ $q'_{cm}$=0 .
In the experiments performed so far, the data for 
$\Delta {\cal M}$ do not exhibit 
any significant dependence in $q'_{cm}$ , so the extrapolation 
to $q'_{cm}$=0 is done by a simple average. This 
is done in bins in the outgoing photon angles $\theta_{CM}$ and $\varphi$,
at fixed $\epsilon$ and $q_{cm}$ . 
The resulting $\Psi_0$ term is then fitted as a linear 
combination of two free parameters, which are 
the structure functions \ $P_{LL}-P_{TT}/\epsilon$ \ and 
\ $P_{LT}$ .

\begin{figure}[htb]
\begin{center}
\begin{minipage}[t]{13 cm}
\epsfig{file=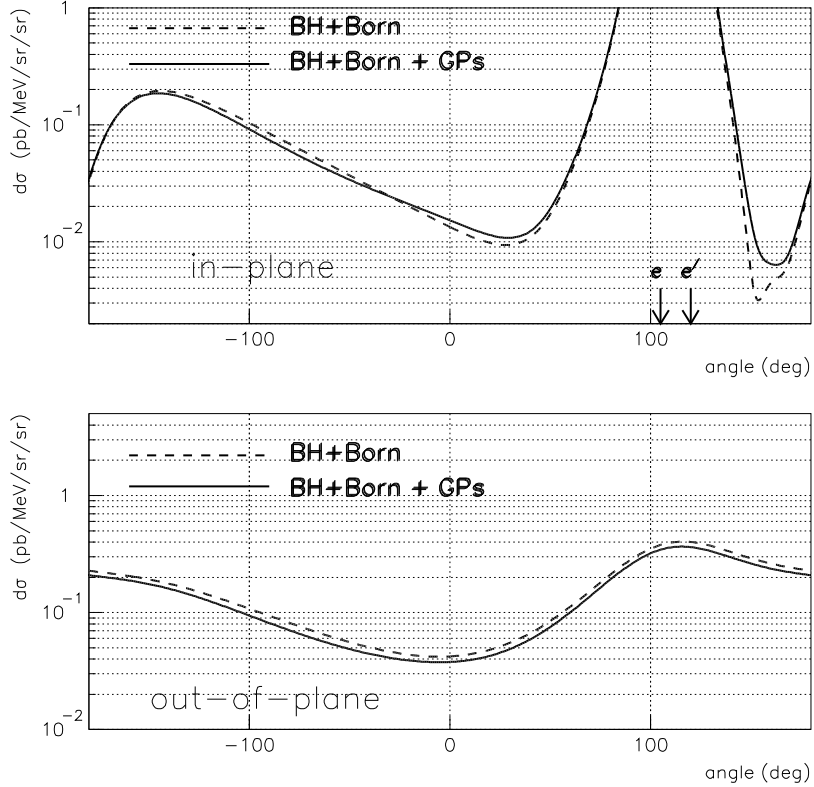,scale=0.6}
\end{minipage}
\begin{minipage}[t]{16.5 cm}
\caption{$(ep \to ep \gamma )$ cross section at $q_{cm}=$ 1080 MeV/c,
$q'_{cm}=$ 105 MeV/c, $\epsilon=$ 0.95, $\varphi=$ 0 or 180$^{\circ}$
(top) or out-of-plane (bottom).
In abscissa is the longitude angle of the outgoing
photon when the polar axis is oriented perpendicular to $\vec q_{cm}$.
Top plot: longitude=0 when $\theta_{CM}=0$. Bottom plot: latitude
= 45$^{\circ}$.  Dashed curve = BH+Born, solid curve = BH+Born + 
a realistic first order GP effect. 
}
\label{fig05}
\end{minipage}
\end{center}
\end{figure}

Figure \ref{fig05} shows an example of the angular dependence
of photon electroproduction cross section, 
for in-plane and out-of-plane kinematics. 
The polarizabilities are responsible for the difference between
the full and dashed curves. In-plane, the pattern of the cross section
and of the polarizability effect is rather complicated, due to the
strong BH-VCS interference. The most suitable region to measure GPs 
is away from the BH peaks (located by the $e$ and $e'$ arrows on the 
graph). Out-of-plane, the cross section behaves much more smoothly 
and the GP effect is almost constant.

\subsection{The Dispersion Relation Model for VCS} \label{secdr}

B.Pasquini et al. developed a model for Real and Virtual
Compton Scattering based on Dispersion Relations~\cite{drmodel1,drmodel2}.
This approach has several advantages, including a rather large
range of validity in $Q^2$, and an applicability in the $\Delta (1232)$
resonance region, where the LET does not hold.
In this model the VCS Non-Born amplitudes are given by
dispersive integrals. They can be split in two parts:
1) a ``$\pi N$'' part which describes the $\pi N$ intermediate states,
and is calculated using MAID pion photoproduction amplitudes~\cite{maid};
2) a part due to asymptotic behavior plus contributions beyond ``$\pi N$''.
The spin GPs are entirely predicted, but not the two scalar GPs. 
The asymptotic behavior of  \ $\alpha_E(Q^2)$ \ and \ $\beta_M(Q^2)$ \
has to be parametrized, and for that purpose a dipole form has been chosen: 

\begin{eqnarray}
\alpha_E(Q^2) - \alpha_E^{\pi N}(Q^2) \ = \  
{ \displaystyle [ \ \alpha_{E}^{exp}   -  
\alpha_{E}^{\pi N} \ ]_{ \ Q^2=0} 
\over
\displaystyle ( \ 1 + Q^2/ \Lambda_{\alpha}^2 \ )^2 }
\label{eq04}
\end{eqnarray}
(same relation for $\beta$ with parameter $\Lambda _{\beta}$). 
 $\alpha_E^{\pi N}$ ($\beta_M^{\pi N}$)
is the $\pi N$ dispersive contribution evaluated from MAID~\cite{maid}, 
 $\alpha_E^{exp}$ ($\beta_M^{exp}$) is the experimental value 
of the polarizability at $Q^2=0$~\cite{olmos}, and the mass coefficients 
$\Lambda _{\alpha}$ and $\Lambda _{\beta}$ are the two 
free parameters to be fitted from experiment. 

In an experimental analysis based on the DR model,
one compares the measured cross section \  $d^5 \sigma _{EXP}$ \ to the 
one predicted by the model for several values
of $\Lambda _{\alpha}$ and $\Lambda _{\beta}$ . 
The fitted values of the parameters are the ones which minimize
the $\chi^2$ between experimental and theoretical cross sections.
\footnote{ Such fits are performed locally in $Q^2$, so the result
is practically insensitive to the somewhat arbitrary dipole form 
chosen for \ $\alpha_E(Q^2) - \alpha_E^{\pi N}(Q^2)$ \ \ and \ \
$\beta_M(Q^2) - \beta_M^{\pi N}(Q^2)$ .}
Then it is straightforward to determine the scalar GPs
by use of eq.~\ref{eq04}, and also if desired
the structure functions \ 
$P_{LL}$ , $P_{TT}$ \ and \ $ P_{LT} $ .

\section{The unpolarized experiments }

Table \ref{tab02} summarizes the three first VCS experiments
which have been dedicated to the determination of GPs, at MAMI, 
the Thomas Jefferson National Accelerator Facility (JLab) and MIT-Bates.
They have measured unpolarized $(ep \to ep \gamma)$ cross sections 
and extracted GPs and structure functions by the methods cited above.

\begin{table}[h]
\begin{center}
\begin{minipage}[t]{16.5 cm}
\caption{The first dedicated VCS experiments.}
\label{tab02}
\end{minipage}
\begin{tabular}{|c|c|c|c|c|c|}
\hline
Lab & $Q^2$ (GeV$^2$) & CM energy W  & $\epsilon$ & data taking
&  status \\
\hline
MAMI-A1 VCS & 0.33   & $< (m_N + m_{\pi})$ & 0.62 &
 1995+97 & publ.2000~\cite{mami1}  \\
JLab E93-050 & 1.0, 1.9 & up to  1.9  GeV & 0.95, 0.88 &
  1998 &  publ.2004~\cite{jlab1,jlab2} \\
Bates \ E97-03 & 0.05    & $< (m_N + m_{\pi})$ & 0.90 &
  2000 &  in analysis~\cite{bates01} \\
\hline
\end{tabular}
\end{center} 
\end{table}

These experiments detect the outgoing electron and proton in 
magnetic spectrometers. High resolution is crucial, since the 
outgoing photon is reconstructed as the missing particle. 
High luminosities are also required, due to the smallness of photon 
electroproduction cross sections. A high duty cycle is necessary
to minimize the rate of accidental coincidences.
Below pion threshold, the elastic process \ $(ep \to ep)$ \ 
is kinematically close to VCS and may induce background in the
detectors.
Finally, absolute cross sections must be determined accurately
in order to extract the small effect due to the polarizabilities.
The acceptance of the apparatus must be calculated with great care by a 
Monte-Carlo simulation~\cite{luc}.
For all these reasons, such experiments are difficult.

The main sources of systematic errors arise from: 
energy and angle calibration of the spectrometers, 
luminosity calculation, uncertainty in the radiative corrections to 
photon electroproduction~\cite{radcor}, the choice of form factors 
in the (BH+Born) cross section, and also possible cross section shape 
departures from fit hypothesis.

\subsection{The experimental results from LEX analyses }

\begin{figure}[htb]
\begin{center}
\begin{minipage}[t]{13 cm}
\epsfig{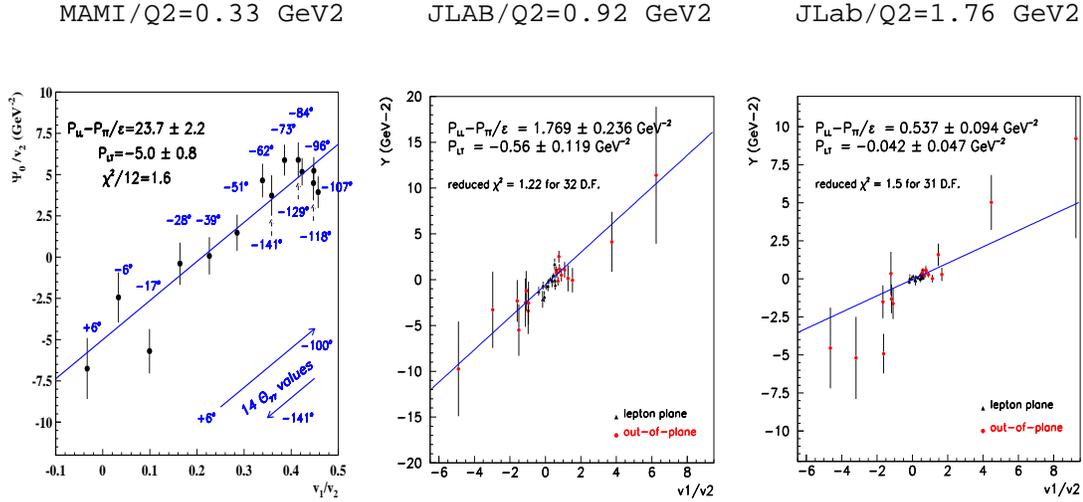}
\end{minipage}
\begin{minipage}[t]{16.5 cm}
\caption{The polarizability fit of the LEX analyses of the 
MAMI and JLab experiments. In ordinate: \ 
$\Psi_0/v_2 = P_{LT} + {v_1 \over v_2} (P_{LL} - P_{TT}/\epsilon)$
 \ (cf.eq.~\ref{eq02}).
\label{fig06}}
\end{minipage}
\end{center}
\end{figure}

Figure \ref{fig06} is a visual representation of the fit 
performed  on the $\Psi_0$ term of eq.~\ref{eq02}, 
in the MAMI~\cite{mami1} and JLab~\cite{jlab1} experiments.
The polarizability signal shows up clearly, in the non-zero
slope $(P_{LL}- P_{TT}/\epsilon)$ and intercept $( P_{LT} )$ 
of the fitted straight line.

The LEX result of MAMI is at low $Q^2$ and therefore has been compared
with most theoretical model predictions of the VCS structure 
functions~\cite{ericedhose}. As a summary,
the values of $(P_{LL}- P_{TT}/\epsilon)$ and  $( P_{LT} )$
measured by this experiment agree well with the 
calculation of heavy baryon chiral perturbation theory
(HBChPT)~\cite{chpt1,chpt2},
while they disagree with all other model predictions.
The HBChPT calculation is done at order $p^3$.
The calculation at next order $p^4$ is not fully available yet,
it is published only for the spin GPs~\cite{chptop4}.

At higher $Q^2$, 
the LEX result of JLab cannot be compared to these
low energy models. The first statement one can deduce from the
numerical values reported in figure~\ref{fig06} is that 
the structure functions definitely show a
strong fall-off with four-momentum transfer. A more complete discussion
is postponed to the next section, in connection with the DR model.

\subsection{The experimental results from DR analyses}

The Jlab VCS data has been analyzed using the DR approach
as explained in section \ref{secdr}. The mass coefficients
$\Lambda _{\alpha}$ and $\Lambda _{\beta}$ which parametrize the 
behavior of the asymptotic contribution to the GPs $\alpha_E$
and $\beta_M$ have been fitted
on experimental photon electroproduction cross sections.
This has been done for three different data sets,
covering different values of $Q^2$ (0.9 and 1.8 GeV$^2$) and W
(below pion threshold and up to 1.3 GeV).
The range of values found is 
[0.70 -0.77] GeV for $\Lambda _{\alpha}$  and
[0.63 -0.79] GeV for $\Lambda _{\beta}$  \cite{jlab1}.
This indicates that the asymptotic contribution to the scalar GPs
falls off more rapidly than the standard dipole elastic form factor
(having $\Lambda=0.84$ GeV).

\begin{figure}[htb]
\begin{center}
\begin{minipage}[t]{13 cm}
\epsfig{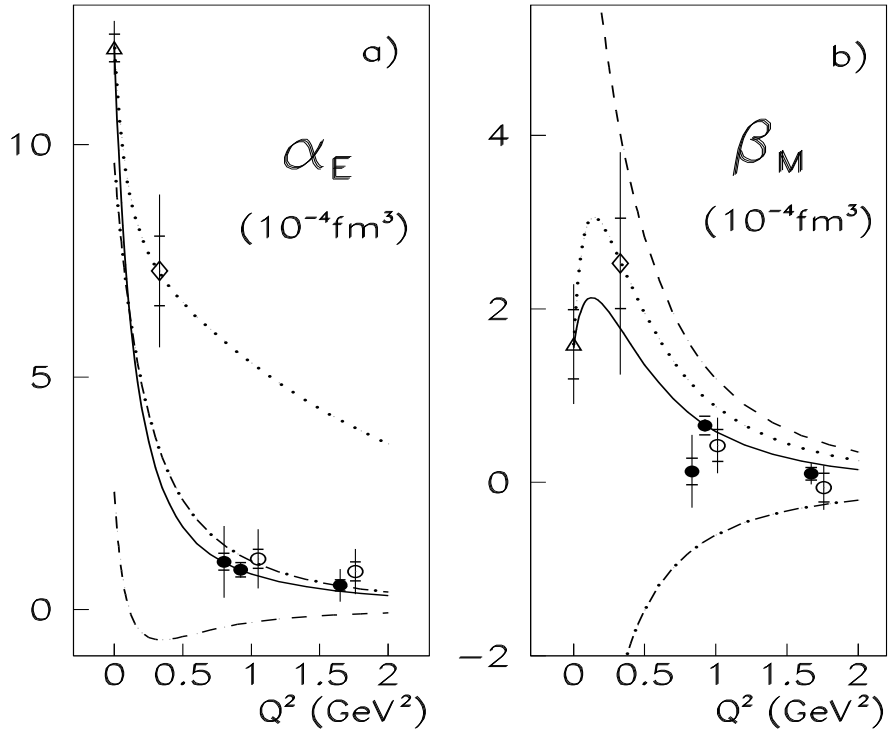}
\end{minipage}
\begin{minipage}[t]{16.5 cm}
\caption{Compilation of the data on electric (a) and
magnetic (b) GPs of the proton. 
Data points at $Q^2$=0 are from Ref.~\cite{olmos}
({\tiny $\bigtriangleup$}). The other points are the LEX analysis of 
MAMI~\cite{mami1} ($\diamond$)  and the
LEX ($\circ$) and DR ($\bullet$) analyses of JLab~\cite{jlab1}.
Some JLab points are shifted in abscissa for better visibility.
The inner error bar is statistical; the outer one is the total
error. The curves show calculations 
in the DR model, as explained in the text.
}
\label{fig07}
\end{minipage}
\end{center}
\end{figure}

In such an analysis the determination of the electric and
magnetic GPs is straightforward, from eq.~\ref{eq04}.
Figure \ref{fig07} shows the world results for these GPs
deduced from the MAMI and JLab experiments, including 
LEX and DR analyses. To make this figure, the LEX results, 
initially in terms of structure functions, have been
translated in terms of the electric and magnetic GPs
by use of eq.~\ref{eq03}. In these equations we took
the spin GPs  as predicted by the DR model, and the proton EM 
form factors were calculated using 
the parametrization of ref.~\cite{brash}.

The curves on figure~\ref{fig07} are calculated using the DR model.
The solid curve is the DR calculation for
$\Lambda_{\alpha}$=0.70 GeV and $\Lambda_{\beta}$=0.63 GeV,
as fitted on one of the JLab data sets. 
The dotted curve is the DR calculation for 
$\Lambda_{\alpha}$=1.79 GeV and $\Lambda_{\beta}$=0.51 GeV,
which reproduces the MAMI LEX point.
By definition (cf. eq.\ref{eq04}) all DR curves are 
constrained to go through the experimental RCS point at $Q^2=0$ .

The fact that there is no unique DR curve going through all 
the data points, especially for the electric polarizability, 
does not invalidate the model. It simply means  
that the dipole parametrization of eq.~\ref{eq04} does not hold 
over the entire range of $Q^2$. Another fact to be aware of is
the model-dependency introduced in this figure by transforming the LEX
result (structure functions) into GPs. For example, the structure
function $P_{TT}$ involved in this operation behaves quite
differently according to different theoretical 
predictions~\cite{drmodel1,chptop3} and has convergence problems
in ChPT~\cite{chptop3,chptop4}. Its separate measurement 
would be of great interest.

In fig.~\ref{fig07} the DR calculation corresponding to the solid curve
has been split into its $\pi N$ part (dashed) and asymptotic part 
(dot-dashed). While the electric polarizability
is dominated by the asymptotic part,
for the magnetic polarizability both contributions are important.
The $\pi N$ contribution to $\beta_M$ is strongly paramagnetic,
predominantly arising from the $\Delta(1232)$ resonance. 
The asymptotic part of $\beta_M$ is strongly diamagnetic and associated
to  $\sigma$-meson $t$-channel exchange. The interplay results in
a turn-over of \ $\beta_M$ \ at small $Q^2$, as already mentioned
in section \ref{secmeaning}. The Bates experiment~\cite{bates01}
performed at $Q^2=0.05$ GeV$^2$ (analysis in progress) will give 
new insight on this specific behavior of  $\beta_M$ .

\subsection{VCS in the resonance region}

\begin{figure}[htb]
\begin{center}
\begin{minipage}[t]{13 cm}
\epsfig{file=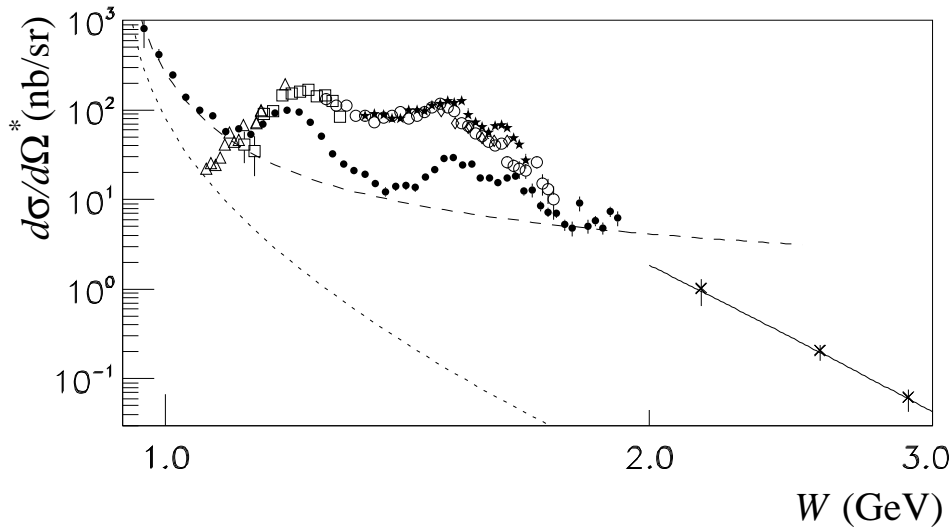,scale=1.4}
\end{minipage}
\begin{minipage}[t]{16.5 cm}
\caption{VCS and RCS in the resonance region.
Photon electroproduction cross section data of 
JLab experiment E93-050 divided by the virtual photon flux 
factor (black dots),
compared to various large angle RCS data:
 ($\star$)~\cite{Wada:1984sh},
($\diamondsuit$)~\cite{Jung:1981wm},
  ($\triangle$)~\cite{Hallin:1993ft},
 ($\circ$)~\cite{Ishii:1985ei},
 (box)~\cite{Wissmann:1999vi},
 ($\times$)~\cite{Shupe:1979vg}).
 The solid curve is an $s^{-6}$ power law normalized to the
 $W=2.55$~GeV RCS Cornell point.
 The dashed curve is the BH+Born+$\pi^0$-exchange 
 cross section and the dotted curve the BH alone.
}
\label{fig08}
\end{minipage}
\end{center}
\end{figure}

In the JLab experiment E93-050,  photon electroproduction
in the resonance region was investigated for the first time,
and cleanly separated from the $(ep \to ep \pi^0$) channel.
Cross sections have been measured up to W=1.9 GeV,
at $Q^2=1$ GeV$^2$ and backward angle $\theta_{CM}$ \cite{jlab2}.
In figure~\ref{fig08} resonant structures are clearly visible
up to W=1.7 GeV. At higher W, this data together with existing 
large-angle RCS data  show a transition to a $Q^2$-independent regime,
suggesting that the VCS process starts coupling to elementary quarks.

\section{Present and Perspectives}

The {\it unpolarized} VCS program can be pursued along several directions:

\vskip 2 mm \par\noindent
- a more detailed mapping of the structure functions
\ $(P_{LL} - P_{TT}/\epsilon)$ \
and \ $ P_{LT} $ \ versus $Q^2$ \newline
- a Rosenbluth-type separation of the structure functions 
$P_{LL}$ and $P_{TT}$ by cross section measurements at 
different $\epsilon$ \newline
- investigate in more detail photon electroproduction in the 
resonance region up to W=2 GeV, in a larger phase space and
with better statistics (there has been only one 
exploratory experiment) \newline
- in the region of the $\Delta$(1232) resonance, use the 
DR formalism to measure the electric and magnetic GPs at high $Q^2$ 
(up to 4 GeV$^2$ in the proposed  experiment of ref.~\cite{prop4gev}).

\vskip 5 mm
Quite obviously, polarization degrees of freedom open up 
new possibilities of investigations in VCS. At present time 
there are two well-identified cases:

\vskip 2 mm
\begin{tabular}{lll}
Single polarization: & 
 $\vec e p \to ep \gamma $ & with longitudinally polarized incoming 
electrons\\
Double polarization & 
 $\vec e p \to e \vec p \gamma $ & by adding the recoil 
proton polarization measurement. \\
\end{tabular}
\vskip 2 mm

Such a {\it polarized} VCS program is under way at Mainz and is
summarized below.

\subsection{VCS with single polarization}

With a longitudinally polarized electron beam one can study the 
beam spin asymmetry or electron single spin asymmetry (SSA)
in the \ $\vec e p \to ep \gamma $ \ process:

\begin{eqnarray}
SSA = {\displaystyle  d^5 \sigma (+)  - d^5 \sigma (-) \over
\displaystyle  d^5 \sigma (+)  + d^5 \sigma (-) } \ = \ 
{\displaystyle \Delta  \sigma \over \displaystyle 2 \cdot d^5 \sigma_{unpol.} }
\label{eq21} 
\end{eqnarray}
where $d^5 \sigma(+)$ (resp. $d^5 \sigma(-)$) is the cross section for incoming 
electrons of helicity $+{1 \over 2}$ (resp. $-{1 \over 2}$).
Rewriting eq.~\ref{eq00} as: \ 
$T_{ep \to ep \gamma} \ = \ T_{BH} \ + \ T_{VCS}$ \ ,
with the Born part now in $T_{VCS}$ , 
the $(ep \to ep \gamma)$ cross section can be expressed as the sum of 
three contributions:

\begin{eqnarray}
d^5 \sigma \ = \vert T_{BH} \vert ^2 +
             \vert T_{VCS} \vert ^2 + 
             2 \mbox{Re} (T_{BH} \cdot T_{VCS})
\ = \ 
\ d^5 \sigma_{BH} \ + \ d^5 \sigma_{VCS} \ + \  
d^5 \sigma_{Interf.BH-VCS}
\label{eq220}
\end{eqnarray}

In the singly polarized case one can write the numerator of the SSA as:
\begin{eqnarray}
\Delta \sigma \ = \ \Delta \sigma_{BH} \ + \ \Delta \sigma_{VCS}
\ + \ \Delta \sigma_{Interf.BH-VCS}
\label{eq22}
\end{eqnarray}

The first term \ $\Delta \sigma_{BH}$ \ is zero. 
For the second term \  $\Delta \sigma_{VCS}$ \ 
we are in a case analogous to the 
usual decomposition of the hadron electroproduction cross section, 
including the fifth response function, i.e.: 
\begin{eqnarray}
d^5 \sigma_{VCS} \ = \ \Gamma_v \cdot {\big [ }
d \sigma_T + \epsilon d \sigma_L +
\sqrt{2 \epsilon (1 + \epsilon) } d \sigma_{LT} \cos \varphi +
\epsilon d \sigma_{TT} \cos (2 \varphi ) + 
h \sqrt{ 2 \epsilon (1 - \epsilon) } d \sigma_{LT}' \sin \varphi
{\big ] }
\label{eq23}
\end{eqnarray}
where $\Gamma_v$ is the virtual photon flux and $h$ the electron
beam polarization.
As shown in ref.~\cite{guikrollschur},  the $\sigma_{LT}$ term
measures the real part of the longitudinal-transverse interference,
whereas  the $\sigma'_{LT}$ term measures its imaginary part, 
i.e. the relative phase between the longitudinal and transverse
VCS helicity amplitudes.
By flipping the helicity of the incoming electron, one isolates
in \ $\Delta \sigma_{VCS}$ \ the fifth response function \ $\sigma'_{LT}$ .
On the other hand, the
third term  of eq.~\ref{eq22} is usually the dominant one,
in other words the SSA is mainly due to the interference
of the real BH+Born amplitude with the imaginary part of the VCS amplitude
\cite{drmodel1}. One should note that this term does not necessarily 
behave like $\sin \varphi$ , contrary to the second term 
$\Delta \sigma_{VCS}$ .
The single spin asymmetry is one case where the interference of VCS
with Bethe-Heitler is not undesirable. Indeed it serves as an 
amplifier of the measured asymmetry; the SSA would be
much smaller in the case of VCS alone.

The SSA is non-zero only for kinematics above pion threshold 
(because one needs an $Im(T_{VCS})$) and for 
out-of-plane values of $\varphi$ .
By measuring this observable,  one will test the models specifically in
their prediction of $Im(T_{VCS})$. Namely, in the DR model
this tests the contribution of the $\pi N$ intermediate states 
(calculated using MAID pion photoproduction multipoles).

\vskip 2 mm
{\it The MAMI single polarization VCS experiment~\cite{mamivcsssa}: } 
data was taken between 2002 and 2004, 
at $Q^2=0.33$ GeV$^2$, W=1.2 GeV, $\epsilon =0.48$
and forward $\theta_{CM}$, with the proton spectrometer tilted out-of-plane.
In these kinematics, the DR model predicts an SSA which is almost 
a pure $\sin \varphi$ , with a sizeable amplitude of about 10 \% 
\cite{drmodel2}. The measurement of this amplitude will be a very 
specific cross-check of the dispersion formalism for VCS.
The analysis of the experiment is in progress.

The experiment will also measure the SSA for the  \ $ep \to ep \pi^0$ \
channel in the same kinematics, allowing further tests of the theory 
related to the fifth response function of pion electroproduction.

The SSA in the $(ep \to ep \gamma)$ channel, as predicted by the DR model
is little sensitive to the GPs for the Mainz kinematics~\cite{drmodel2}. 
However one can analyze the experiment by doing a helicity sum 
and performing an unpolarized analysis of the DR type, 
as explained in section \ref{secdr}, to extract the electric
and magnetic GPs.

\vskip 2 mm
{\it Link with VCS at higher energy: }
The SSA in exclusive photon electroproduction at higher energy 
has been studied in ref.~\cite{guikrollschur}. The SSA in the 
Deep VCS regime (DVCS) has been studied in numerous papers
\cite{diehl97}.
It is worth recalling that in DVCS this single spin asymmetry 
or its numerator \ $\Delta  \sigma$ \ is one of the main tools 
to investigate the Generalized Parton Distributions (GPDs).
\footnote{ in DVCS experiments, \ $\Delta  \sigma$ \ is 
fitted to the form: \ $A \sin \varphi + B \sin 2 \varphi$ \ 
with \ $A \gg B$. The full theoretical expression of  $\Delta \sigma$
contains a third term  ($ C \sin 3 \varphi $) \cite{diehlthesis}.
}

\subsection{VCS with double polarization}

This is an ideal case, since the process 
 \ $\vec e p \to e \vec p \gamma$ \ 
allows in principle to disentangle the six lowest order GPs.
The full formalism of double polarization observables below pion threshold
has been worked out in ref.~\cite{vdhdoublepol}. One uses a longitudinally
polarized electron beam and measures the polarization
components $P_x, P_y, P_z$ of the recoil proton in the 
$\gamma p$ center-of-mass.
The double spin asymmetry along axis $i$ ($i=x,y,z$) is given by:

\begin{eqnarray}
DSA_{(i)} \ = \ { 
\displaystyle
[ \ d^5 \sigma_{h=+1/2, s'_i \uparrow} \ - \  
    d^5 \sigma_{h=+1/2, s'_i \downarrow} \ ] \ - \
[ \ d^5 \sigma_{h=-1/2, s'_i \uparrow} \ - \  
    d^5 \sigma_{h=-1/2, s'_i \downarrow} \ ]
\over
\displaystyle
[ \ d^5 \sigma_{h=+1/2, s'_i \uparrow} \ + \  
    d^5 \sigma_{h=+1/2, s'_i \downarrow} \ ] \ + \
[ \ d^5 \sigma_{h=-1/2, s'_i \uparrow} \ + \  
    d^5 \sigma_{h=-1/2, s'_i \downarrow} \ ]  
}
\label{eq31}
\end{eqnarray}
where $h$ is the incoming electron helicity and $s'_i$  the 
projection of the recoil proton's spin along axis $i$.
This formula is equivalent to the following definition
of the polarization component $P_i$  of
the recoil proton along axis $i$,
for a given helicity state $h$ of the incident electron:

\begin{eqnarray}
P_i \ = \ {
\displaystyle
d^5 \sigma_{h,s'_i \uparrow} \ - \ d^5 \sigma_{h,s'_i \downarrow}
\over
\displaystyle
2 d^5 \sigma_{unpol.}
} 
\ = \ {
\displaystyle
\Delta \sigma_{h,i}
\over
\displaystyle
2 d^5 \sigma_{unpol.}
} 
\label{eq32}
\end{eqnarray}

The polarization components $P_i$ receive contributions 
from (BH + VCS Born) and (VCS Non-Born),
and similarly to the unpolarized case a Low Energy Theorem can be 
established. This yields the following equations:

\begin{eqnarray}
\Delta \sigma_{h,i} \ = \ \Delta \sigma_{h,i}^{BH+Born} \ + 
 \phi  \cdot q'_{cm} \cdot \Delta \Psi_{0(h,i)} \ + \ {\cal O}(q'^2_{cm})  
\label{eq33}
\end{eqnarray}
where the first term is entirely calculable, and the second term
\ $\Delta \Psi_{0(h,i)}$ is a linear combination of the 
following structure functions~\cite{guivdh98} :

\begin{eqnarray}
\begin{array}{lll}
\Delta \Psi_{0(h,z)} & = & 4 h \times [ \ 
c_1^z \cdot P_{TT} + c_2^z \cdot P_{LT}^z + c_3^z \cdot P'^z_{LT} \ ] \\
\Delta \Psi_{0(h,x)} & = & 4 h \times [ \ 
c_1^x \cdot P^{\perp}_{LT} + c_2^x \cdot P^{\perp}_{TT} + 
c_3^x \cdot P'^{\perp}_{TT} + c_4^x \cdot P'^{\perp}_{LT} \ ] \\
\Delta \Psi_{0(h,y)} & = & 4 h \times [ \ 
c_1^y \cdot P^{\perp}_{LT} + c_2^y \cdot P^{\perp}_{TT} + 
c_3^y \cdot P'^{\perp}_{TT} + c_4^y \cdot P'^{\perp}_{LT} \ ] \\
\label{eq34}
\end{array}
\end{eqnarray}

The $c_n^i$ are kinematic coefficients. The structure functions
of eq.~\ref{eq34}
are themselves linear combinations of the six lowest order GPs. 
Apart from the formulas already given in eq.~\ref{eq03}, the new 
expressions for the doubly polarized case are~\cite{guivdh98} :

\begin{eqnarray}
\begin{array}{lll}
P_{LT}^z \ \ \  & = & {3 \ {\tilde Q} \ q_{cm} \over 2 \ {\tilde q _0} } 
\ G_M \  P^{(01,01)1}  (q_{cm}) -
 {3 \ M_N \ q_{cm} \over  {\tilde Q} } \  G_E \
 P^{(11,11)1}  (q_{cm}) \\
P'^z_{LT} & = & 
- {3 \over 2} \ {\tilde Q} \  
\ G_M \  P^{(01,01)1}  (q_{cm}) +
 {3 \ M_N \ q_{cm}^2 \over  {\tilde Q} \ {\tilde q _0} } \  G_E \
 P^{(11,11)1}  (q_{cm}) \\
P^{\perp}_{LT}  & = & 
 {R \over 2} \ { G_E  \over  G_M } 
\ P_{TT} (q_{cm}) - 
{1 \over 2R} \  { G_M  \over  G_E } \
\ P_{LL} (q_{cm}) \\
P^{\perp}_{TT}  & = & 
- {q_{cm} \over 2 } \ G_M \ 
{\big [ } \ 
3  P^{(11,11)1}  (q_{cm}) + \sqrt{ {3 \over 2} }  P^{(11,11)0}  (q_{cm}) 
\ {\big ]} \\
P'^{\perp}_{TT} & = & 
{q_{cm} \over 2 } \ G_M \ 
{\big [ } \ 
3  { q_{cm} \over {\tilde q _0} } P^{(11,11)1}  (q_{cm}) + 
\sqrt{ {3 \over 2} } \ { {\tilde q _0} \over q_{cm} }\ P^{(11,11)0}  (q_{cm}) 
\ {\big ]} \\
P'^{\perp}_{LT} & = & {3 \ {\tilde Q} \ q_{cm} \over 2 \ {\tilde q _0} } 
\ G_M \  
{\big [ } \ 
P^{(01,01)1}  (q_{cm}) -
\sqrt{ {3 \over 2 } } \ {\tilde q _0} \  
 P^{(11,02)1}  (q_{cm})
\ {\big ]} \\
\label{eq36}
\end{array}
\end{eqnarray}
where $R=2 M_N/ \tilde Q$ .
The GPs can be extracted from the linear systems above, provided
the $P_i$'s are measured over a large range in $\theta_{CM}$ (to allow
enough variation of the  coefficients $c_n^i$).
The structure function $P'^{\perp}_{LT}$ can only be extracted by 
an out-of-plane measurement.

The BH+Born process yields large double polarization asymmetries. 
But the contribution of the GPs to these asymmetries is small,
typically of a few percent, so it is tough to measure.
A recent paper~\cite{kaodoublepol} reviews theoretical predictions 
for these double polarization asymmetries.

At MAMI the first test runs for such a second generation VCS 
experiment~\cite{mamivcsssa} took place in 2004, using the 
focal plane polarimeter and the longitudinally polarized beam.
It's clearly a very challenging experiment, requiring high statistics
and very reduced systematic errors.

\subsection{A Related Topic}

One can do Compton physics on the nucleon without doing an explicit Compton 
scattering experiment. For example in inclusive electron scattering
\ $N(e,e')X$ , one makes an extensive use of the optical theorem that 
relates the cross section of the process to the forward Compton 
scattering amplitude:

\begin{eqnarray}
\sigma (\ \gamma^*({\underline q}) N \ \to \ X \ ) \ \sim \ \mbox{Im} \ T \ 
( \ \gamma^*({\underline q}) \ N \ \to \ \gamma^*({\underline q}) N \ ) 
\end{eqnarray}

Here T is the forward doubly virtual Compton amplitude 
(VVCS), in which the initial and final virtual photons 
have the same four-momentum ${\underline q}$ .

Most of the sum rules established in real photon processes
lead to interesting results by generalization to a virtual photon.
Some of them concern Generalized Polarizabilities; 
one example is the measurement of the Generalized Forward Spin GPs
\ $\gamma_0$ \ and \ $\delta_{LT}$ \ of the neutron~\cite{jpchenfwd}.
This is part of the so-called ``extended GDH program'' performed by
scattering polarized electrons on a polarized Helium3 target:
\ $ \vec { ^3 He} ( \vec e , e' )X$ .

The forward spin polarizability $\gamma_0$ is defined in 
real photon processes by:

\begin{eqnarray}
\gamma_0 \ = \ { \displaystyle 1 \over 2  \displaystyle\pi ^2 } \ 
\int _{\nu_0} ^ { \infty}  \ 
\sigma_{TT} \ {  \displaystyle d \nu \over  \displaystyle \nu ^3 }  
\end{eqnarray}
where $\nu$ is the incident photon lab energy,  $\nu_0$ this energy at
the pion threshold, and \ 
$\sigma_{TT} = -{1 \over 2} (\sigma_{3/2} - \sigma_{1/2}) $ \ 
the difference of photon absorption helicity cross sections.
When going from a real to a virtual photon, one gets a
generalized sum rule for $\gamma_0$ , plus a sum rule defined only 
in the case of a virtual photon for the
longitudinal-transverse spin GP \ $\delta_{LT}$ :

\begin{eqnarray}
\begin{array}{lll}
\gamma_0(Q^2) & = & { \displaystyle 1 \over 2  \displaystyle\pi ^2 } \ 
{\displaystyle \int} _{\nu_0} ^ { \infty}  \ 
{  \displaystyle K(\nu, Q^2) \over \displaystyle \nu} \ 
\sigma_{TT}(Q^2) \ {  \displaystyle d \nu \over  \displaystyle \nu ^3 }  
\\
\ & \ & \\
\delta_{LT}(Q^2) & = & { \displaystyle 1 \over 2  \displaystyle\pi ^2 } \ 
{\displaystyle \int} _{\nu_0} ^ { \infty}  \ 
{  \displaystyle K(\nu, Q^2) \over \displaystyle \nu} \ 
\sigma_{LT}(Q^2) \ 
{  \displaystyle 1 \over Q }
{  \displaystyle d \nu \over  \displaystyle \nu ^2 }  
\\
\end{array}
\end{eqnarray}

$K$ is the virtual photon flux factor.
These new observables have been measured up to $Q^2=$ 0.9 GeV$^2$
\cite{jpchenfwd}
and compared to various ChPT predictions~\cite{kao3,bernard}
and the MAID model~\cite{maid2}.
Contrary to expectations, ChPT has difficulties reproducing 
the forward spin GP \ $\delta_0$ \ even at very low $Q^2$ (0.1 GeV$^2$).
This data represent the first actual measurement of Generalized 
Spin Polarizabilities of the nucleon (here the neutron).
Along the same lines there are results for the 
Generalized Baldin sum rule on \  
$\alpha_E(Q^2) + \beta_M(Q^2)$ \cite{baldinsr}.

It must be emphasized that the Generalized Polarizabilities 
of this section are \ {\it not} \ the same as the ones introduced in the
previous sections.
In VCS we have only one virtual photon, whereas in VVCS (this section)
we have two virtual photons, with identical virtuality. 
These  two types of polarizabilities are however connected 
in the limit $Q^2 \to 0$.

\section{Conclusion}

A nice overview of the history of VCS on the nucleon 
can be found in ref.~\cite{guivdh98}.
It started as a rather unwanted contribution
to radiative corrections to e-N scattering
(proton bremsstrahlung is a VCS Born term).
It regained interest in the 1990's with the hard scattering picture
of QCD~\cite{farrar,pegasys}.
Two very innovative concepts in nucleon structure, 
the Generalized Polarizabilities~\cite{gui95}
and the Generalized Parton Distributions~\cite{ji97}
have lead to dedicated experiments in the last decade.
These experiments, both at low energy (GPs) and high energy (GPDs),  
make full use of the performant electron accelerators
and detectors, and more experimental results are 
expected in the near future. 
The theoretical front is very active too, e.g. in predicting
the Generalized Polarizabilities. Recently, (double) VCS has become
a rather wanted term of radiative corrections to e-N scattering,
since it is proposed to explain the discrepancy between
proton form factors measured by Rosenbluth and polarization transfer
techniques~\cite{gui2gam,blunden}.
To conclude, two-photon physics seems to have a promising future in the 
study of nucleon structure.

\vskip 5 mm
\section*{Acknowledgments }

I wish to thank all my VCS colleagues of the experimental and 
theoretical sides for their help and support, with
special thanks to B. Pasquini, P. Guichon and H. Merkel 
for their proofreading of the manuscript.
The  work done within the JLab Hall A and the MAMI-A1 
collaborations has provided the material of this talk.

This work was supported by DOE, NSF, by contract DE-AC05-84ER40150
under which the Southeastern Universities Research Association
(SURA) operates the Thomas Jefferson National Accelerator Facility
for DOE, by the French CEA and CNRS-IN2P3,
the FWO-Flanders and the BOF-Gent University (Belgium) and by the
European Commission ERB FMRX-CT96-0008.



\end{document}